\documentclass[11pt,a4paper]{article}
\usepackage[left=2.5cm,right=2.5cm]{geometry}
\usepackage{amsmath}
\usepackage{amsfonts}
\usepackage{amssymb}
\usepackage{graphicx, rotating}
\usepackage{epstopdf}
\usepackage{epsfig}
\usepackage{latexsym}
\usepackage{graphicx}
\usepackage{color}
\usepackage{amsmath,amssymb}
\usepackage{slashed}
\usepackage{hyperref}
\usepackage{booktabs}
\usepackage{multirow}
\usepackage{hyperref}
\usepackage{bbm}
\usepackage{lscape}
\usepackage{amscd}

\allowdisplaybreaks

\begin{document}

\begin{titlepage}
\begin{flushright}
\end{flushright}
\vspace{.3in}

\begin{center}
\vspace{1cm}

{\Large \bf
Observing nonstandard $W'$ and $Z'$ through the third generation and Higgs lens
}

\vspace{1.2cm}
{\large Lisa Edelh\"auser and Alexander Knochel}
\vspace{.8cm}

{\it {Institut f\"ur Theoretische Teilchenphysik und Kosmologie, RWTH Aachen,
Aachen, Germany}}\\

\begin{abstract}
\medskip
\noindent
We reinterpret $W'$, $Z'$ and Higgs searches at the LHC in terms of a model
with nonuniversal $V'$ couplings to fermions in order to gauge how well these
searches can be applied outside the (simplified) scenarios for which they were
optimized.  In particular, we consider bounds from $V'$ searches in final
states $\tau \tau$, $e\nu$, $ll$ and $t b$, and discuss the impact of width effects.
We then show that decays of the type $V' \rightarrow V h$ yield additional
bounds on the heavy vector masses and, in the case of a discovery, provide an
important probe of the heavy sector. Finally, we match the low energy limit of
the model to an effective theory and compare the bounds on the resulting
dimension-6 operators with the direct searches discussed in this paper.
\end{abstract}
\vspace{.4cm}

\end{center}
\vspace{.8cm}


\end{titlepage}



\section{Introduction}
Heavy vectors, either in the form of new gauge bosons or as composite
resonances, are a feature of many BSM scenarios such as composite Higgs models,
GUTs with extra $U(1)$ gauge groups, $LR$-symmetric models or models with extra
dimensions. Searches at the LHC for ``classical'' heavy gauge vectors often
consider the sequential Standard model (SSM) case where a $W'$ or $Z'$ are
present with a large mass but couplings identical to the known electroweak
bosons. The SSM-like $W'$ with righthanded couplings allows for a somewhat
simplified analysis as interference with the SM $W$ is eliminated.  In the case
of the $Z'$, there are some typical anomaly-free candidate charge assignments
inspired e.g. from $B-L$ symmetry or $E_6$ GUTs which differ somewhat from the
SSM case, and are subject to RG running \cite{Braam:2011xh,Rizzo:2012rf}.

The family $SU(2)_1 \times SU(2)_2$ model \cite{Chiang:2009kb} on which we
focus here is another possible setting for heavy vector resonances beyond the
SSM case, and it contains an extended Higgs sector charged under the extended
gauge group. The model is of interest for several reasons: it is being used by
CMS as a theory motivation and testing ground for upcoming searches involving
third generation fermions. It provides a concrete realization of enhanced third
generation couplings to the heavy sector. Since family- nonuniversal gauge
groups are strongly constrained from flavor physics, it is useful to know when
the reach of the LHC can beat indirect constraints in such models.
Furthermore, while featuring a SM-like light Higgs by design, as we point out,
the structure of the extended Higgs sector is revealed through interactions
with the heavy gauge bosons even when the physical heavy Higgs bosons are out
of reach of the LHC. We use the opportunity to compare the description of
anomalous Higgs interactions via dimension-6 operators to the low-energy limit
of an explicit model, and compare the sensitivity of the two approaches and its
dependence on the underlying physics.

This paper is organized as follows: 
In section \ref{sec:model} we briefly describe the model given in
\cite{Chiang:2009kb}, and provide the Feynman rules needed for our analyses. We
discuss in particular the (heavy) Higgs sector and its influence on the $V'$
phenomenology. In section \ref{sec:searches}, we revisit existing LHC searches
for heavy vectors, apply them to the $SU(2)_1\times SU(2)_2$ model where
possible and discuss the issues with the interpretation of shape-based versus
single-bin analyses. In section \ref{sec:higgs} we repurpose existing Higgs
searches in $Vh$ associated production final states and derive bounds on $V'$
from $V' \rightarrow V h$ decays. We briefly discuss simple ways how these
searches can be optimized to greatly enhance their reach.  Finally, we contrast
direct searches in the Higgs channel with indirect bounds on EFTs. We conclude
in section \ref{sec:conclusions}.

\section{The Model}
\label{sec:model}
The model which we consider is identical to the one introduced in
\cite{Chiang:2009kb}. We only briefly revisit the basics and some aspects of
particular importance to this work, and refer the reader to the original paper
for a more thorough discussion.  There is an $SU(2)_1\times SU(2)_2\times
U(1)_Y$ electroweak gauge group with couplings $g_1=g/\cos(\theta_E),
g_2=g/\sin(\theta_E), g'$ which is broken down to $SU(2)_L\times U(1)_Y$ with
couplings $g, g'$ by a bifundamental scalar $\eta \sim {\bf (2,2)_0}$. The
latter can be chosen to be selfdual in the sense that
$\epsilon_{\beta\alpha}\epsilon_{\beta'\alpha'}\eta^{*\alpha\alpha'}=\eta_{\beta\beta'}$,
thus eliminating all degrees of freedom besides a (heavy) neutral Higgs scalar
$h'$ and three would-be Goldstone bosons for the heavy vectors $X^{0,\pm}$:
\begin{equation}
\eta = \left[\begin{matrix}u + \frac{h' + i X^0}{2} & \frac{X^+}{\sqrt{2}} \\ -\frac{X^-}{\sqrt{2}} &u + \frac{h' - i X^0}{2}  \end{matrix}\right]
\end{equation}
where we have assumed that $\eta$ develops a vev $\langle \eta\rangle = u
I_{2\times 2}$, and normalize the fields to a kinetic term
$Tr[D_\mu\eta^\dagger D^\mu\eta]$.  There are two Higgs doublets $\phi_1 \sim
{\bf (2,1)_{\frac12}}$ and $\phi_2 \sim {\bf (1,2)_{\frac12}}$
charged under $SU(2)_1$ and $SU(2)_2$ respectively. In analogy to other 2HDMs,
they are set up to have vevs $v_1$ and $v_2$ where as usual, $\tan\beta =
v_2/v_1$.  The authors of \cite{Chiang:2009kb} introduce a small
quantity\footnote{Note that  we normalize the vev differently, with $\langle
\phi_i^\dagger\phi_i^{\phantom{\dagger}} \rangle=v^2_i/2$.} $\epsilon =
v/\sqrt{2}u$ which allows to expand effects in leading powers of $\epsilon$.
Unlike standard 2HDMs, here $\phi_1$ bestows masses to the first two
generations, whereas $\phi_2$ does the same for the third generation. Hence, a
large $\tan\beta \gg 1$ can explain the
relative smallness of masses in the first two generations compared to the top,
whereas it does not explain the hierarchy $m_b \ll m_t$. This is in contrast to
the MSSM or Type II models where $\tan\beta \gg 1$ can explain the hierarchy
$m_b \ll m_t$, but not $m_u, m_c \ll m_t$.  This division of labor for the
$\phi_i$ requires the left handed fermions from the first two generations to be
doublets under $SU(2)_1$, whereas those in the third are doublets under
$SU(2)_2$.  Consequently, the covariant derivatives for the matter and Higgs
fields are given by
\begin{eqnarray}
D_\mu \phi_i &=&\left[\partial_\mu - i g_i W_{i\mu}^a T^a  - i g' \frac12 B_\mu  \right]\phi_i\nonumber \\
D_\mu \psi_{gen=1,2} &=&\left[\partial_\mu - i g_1 W_{1\mu}^a T^a P_L  - i g' Y_f  B_\mu  \right]\psi_{gen}\nonumber \\
D_\mu \psi_{gen=3} &=&\left[\partial_\mu - i g_2 W_{2\mu}^a T^a P_L - i g' Y_f  B_\mu  \right]\psi_{gen}\nonumber \\
D_\mu \eta &=&\partial_\mu\eta - i g_1 W_{1\mu}^a T^a \eta + i g_2 W_{2\mu}^a \eta T^a\,.
\end{eqnarray}
When $\eta$ develops the vev, only transformations $U_1 \eta U_2^\dagger$ with
$U_1=U_2=U$ survive, where $U$ can be identified as elements of the SM group
$SU(2)_L$. Due to nonabelian gauge invariance, this remaining light $SU(2)_L$
couples universally to all generations.  Neglecting $\mathcal O(\epsilon)$
mixing effects, the light and heavy gauge bosons $W_L, W_H$ are given by
\begin{equation}
W^a_1=\frac{g_2 W^a_L + g_1 W^a_H}{\sqrt{g_1^2+g_2^2}},\quad W^a_2=\frac{g_1 W^a_L - g_2 W^a_H}{\sqrt{g_1^2+g_2^2}}\,.
\end{equation}
This approximation is sufficient for our purposes, since $\mathcal
O(\epsilon^2)$ effects such as $V'\rightarrow VV$ decays are subleading in our
analyses. The $W^3_L$ mixes with $B$ to form $\gamma,Z$ as usual.

The particle content of the two doublets is that of usual 2HDMs, and since we
assume to be in the decoupling limit, pseudoscalar, charged and neutral Higgs
mixing are all given by $\beta$,
\begin{eqnarray}
\phi_1&=&\left(\begin{matrix} c_\beta G^+ -s_\beta H^+ \\ \frac{c_\beta}{\sqrt{2}}(v+h+i G^0)-\frac{s_\beta}{\sqrt{2}}(H+i A) \end{matrix}\right)\nonumber\\
\phi_2&=&\left(\begin{matrix} s_\beta G^+ + c_\beta H^+ \\ \frac{s_\beta}{\sqrt{2}}(v+h+i G^0)+\frac{c_\beta}{\sqrt{2}}(H+i A) \end{matrix}\right)\,.
\end{eqnarray} 
By definition, $\beta$ is chosen such that $G^+$, $G^0$ are the would-be
Goldstone bosons of $W,Z$.  In the full theory, $H^+$ and $A$ are not purely
the physical heavy Higgs, but mix with $X^{0\pm}$ to form the heavy sector
would-be Goldstones. These mixings are however $\propto c_\beta
s_\beta=s_\beta^2/\tan_\beta$ and will not concern us for the remainder of this
work where we consider large $\tan_\beta$. Indeed, in the limit where
$m_{A,H,H^+}> m_{V'}$, the interactions relevant for $V'$ searches all
originate from the kinetic terms
\begin{equation}
\mathcal L_{kin} = \sum_f \overline \psi_f D\hspace{-1.5ex}/\, \psi_f + \sum_i D_\mu \phi_i^\dagger D^\mu \phi_i^{\phantom{\dagger}}
\end{equation}
while decays of $V'$ to heavy Higgs particles are additionally governed by
\begin{equation}
\mathcal L_{kin} = Tr[D_\mu \eta^\dagger D^\mu \eta]\,.
\end{equation}
One interaction which has not been considered in
\cite{Chiang:2009kb,Kim:2014afa,Kim:2011qk} is the $V'Vh$ vertex which will be
of particular interest later. For some regions of parameter space, it increases
the width of the heavy vectors by up to $6\%$, and it yields an interesting
phenomenology.  
\subsection{Feynman Rules for Production and Decay}
The couplings of $W'$ and $Z'$ are purely left-handed in our approximation
suitable for collider searches where we neglect $\mathcal O(\epsilon)$ $Z-Z'$
mixing. Hence, while the $W'$ can be SSM-like, the $Z'$ is essentially
$W'$-like and the three states form a near-degenerate $SU(2)_L$ triplet. This
is also reflected in an identical total width.  We are here mainly concerned
with the vertices of $V'$ to SM fermions as well as the $V'Vh$ vertices.  They
are given in Table \ref{feynmanrules}.  
\begin{table}
\begin{center}
\begin{tabular}{|c|c|}
\hline
$ h W^+ {W'}^-, h W^- {W'}^+$ & $\displaystyle i \frac{g^2 v}{2} \frac{s_E^2 -s_\beta^2}{c_E s_E} g_{\mu\nu} $ \\ \hline
$h Z Z'$ & $\displaystyle i \frac{g^2 v}{c_W^2} \frac{s_E^2 -s_\beta^2}{c_E s_E} g_{\mu\nu}$ \\ \hline
$\overline b b Z',\overline \tau \tau Z'\quad /\quad\overline t t Z',\overline \nu_\tau \nu_\tau Z' $ & $\displaystyle \pm\frac{i g}{2} \cot_E \gamma^\mu P_L$ \\ \hline
$\overline b t {W'}^-, \overline t b {W'}^+, \overline \tau \nu_\tau {W'}^-, \overline \nu_\tau \tau {W'}^+ $ & $\displaystyle - \frac{i g}{\sqrt{2}} \cot_E \gamma^\mu P_L $ \\ \hline
$\overline u u Z', \overline \nu_l \nu_l Z' \quad /\quad \overline d d Z', \overline l l Z'$ & $\displaystyle \pm\frac{i g}{2} \tan_E \gamma^\mu P_L$ \\ \hline
$\overline u d {W'}^+, \overline \nu_l l {W'}^+$ & $\displaystyle  \frac{i g}{\sqrt{2}} \tan_E V \gamma^\mu P_L$  \\ \hline
$\overline d u {W'}^-, \overline l \nu_l {W'}^-$ & $\displaystyle \frac{i g}{\sqrt{2}} \tan_E V^\dagger\gamma^\mu P_L $ \\ \hline
\end{tabular}
\end{center}
\caption{The Feynman rules relevant for $V'$ production and decay in the
family-$SU(2)_1\times SU(2)_2$ model. We assume that the heavy Higgs sector has
masses $m_H,~m_{h'}\gg m_{V'}$.\label{feynmanrules}}
\end{table}
As expected, the $Z', W'$ simply couple like an $SU(2)$ triplet. One finds that
the $W'$ couplings become SSM-like (up to a sign in the third generation!) for
$\tan_E=\cot_E=1$ corresponding to $c_E=s_E=1/\sqrt{2}$.  For $\tan_\beta \gg
1$, the factor $s_E^2-s_\beta^2$ in the $V'Vh$ coupling lets the Higgs and
Goldstones behave like a partial fermion of the third generation because
$s_E^2-s_\beta^2 \longrightarrow - c_E^2$, and as one of the first and second
generation for $\tan_\beta \ll 1$. In fact, the decays  $V'\rightarrow V h$ are
dominated by the longitudinal mode and can be approximated by $V' \rightarrow
G^{0\pm} h$. In the limit $\tan_\beta \gg 1$, $BR(Z'\rightarrow Z h)=\frac12
BR(Z'\rightarrow \overline\tau \tau)$ and $BR(W' \rightarrow W h)=\frac14 BR(W'
\rightarrow \tau \nu_\tau)$ if we neglect phase space factors. Consequently,
observing these decays of the heavy vectors gives us information about the
relative mixing angles in the Higgs and gauge sector. Since the $SU(2)$ triplet
couples universally to all fermions of a generation, we can calculate the
widths and branching ratios at tree level simply by counting degrees of
freedom.  Neglecting phase space factors, in the limit $\tan\beta \gg1$,
\begin{equation}
\Gamma_{W'} = \Gamma_{Z'} = \Gamma_{W'}^{SSM}\times \frac{(3+1+\frac14)\cot_E^2 + (6+2)\tan_E^2}{12+\frac14}\,.
\end{equation}
Similarly, the branching fractions scale like $BR = BR_{SSM}\times \cot^2_E
\Gamma_{SSM}/\Gamma_{W'}$ for 3rd generation fermions and $BR = BR_{SSM}\times
\tan^2_E \Gamma_{W'}^{SSM}/\Gamma_{W'}$ for 1st and 2nd generation fermions.
The $\overline q q$ initiated production at LHC scales with $\tan_E^2$, while
the branching into 3rd generation fermions scales as $\cot_E^2$. Neglecting
phase space factors, the $V'$ production  and decay into 3rd generation
fermions thus scales as $\sigma\times BR \propto
\Gamma_{W'}^{SSM}/\Gamma_{V'}$. The $V'$ production with subsequent decay into
1st and 2nd generation fermions receives an additional factor $\sigma\times BR
\propto \tan_E^4 \Gamma_{W'}^{SSM}/\Gamma_{V'}$.  For $\tan_E>1$, the
production and decay into 1st and 2nd generation fermions is thus greatly
enhanced, whereas for $\cot_E>1$, decays into 3rd generation fermions become
important.  \label{sec:frules}

\section{Limits from direct $Z'$ and $W'$ Searches}
\begin{figure}[htp]
\centering
\includegraphics[height=7.5cm]{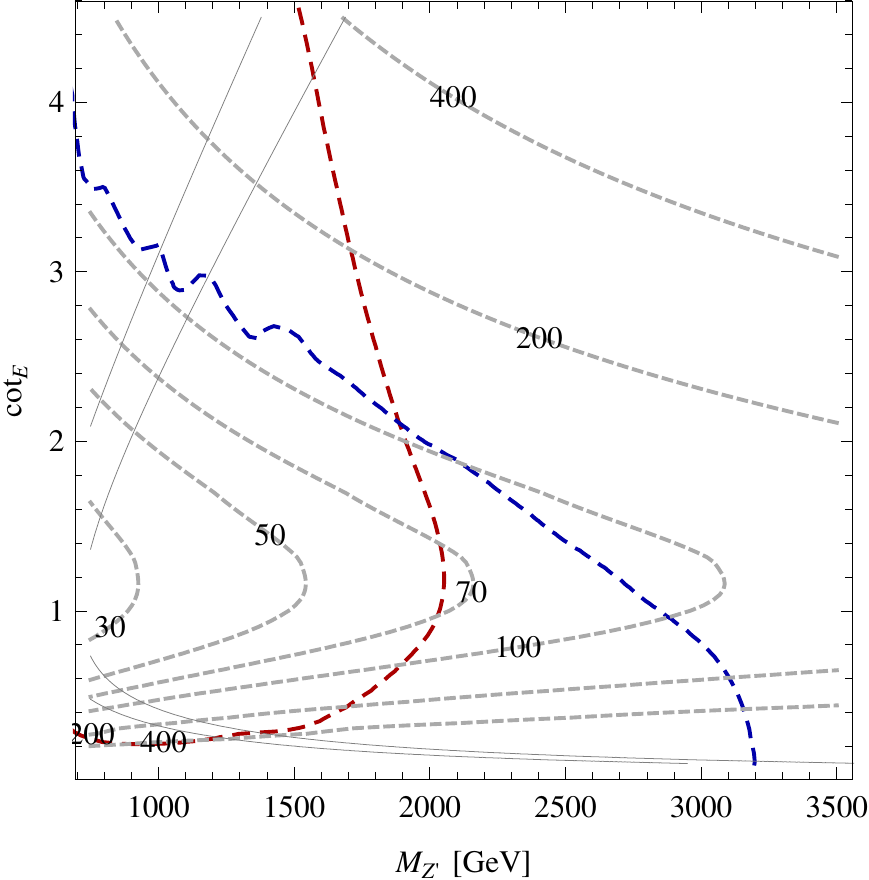}
\caption{Limits from $W'\rightarrow tb$ (red dashed) \cite{Chatrchyan:2014koa},
$Z'\rightarrow ll$ (blue dashed) \cite{Aad:2014cka} from naive scaling. The
gray dashed lines give the width of the W'/Z' in GeV. The gray solid lines
denote the regions where $\epsilon^2<0.15,0.1$ (from left to
right)\label{fig:searches2}} \end{figure}

\begin{figure}[htp]
\raisebox{0.3ex}{\includegraphics[height=7.5cm]{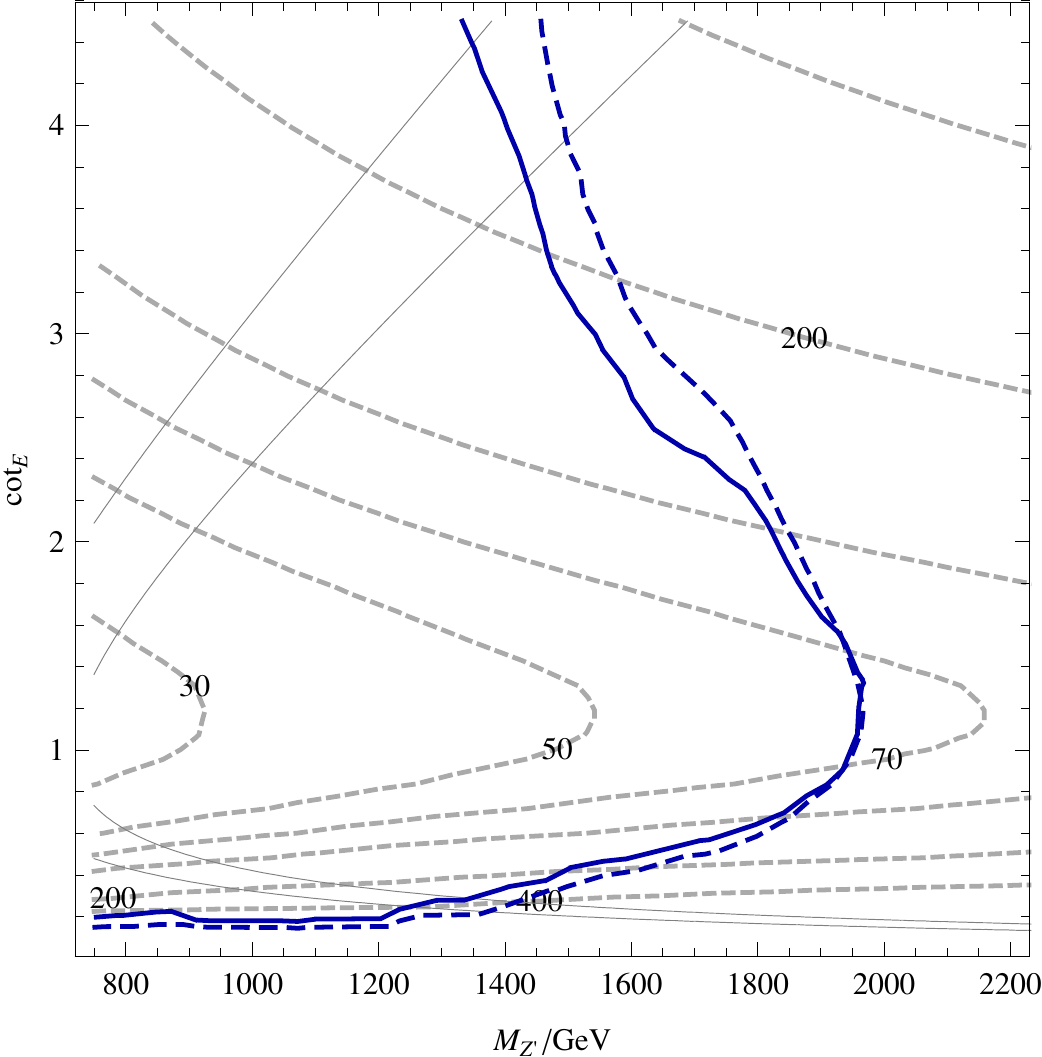}}\includegraphics[height=7.55cm]{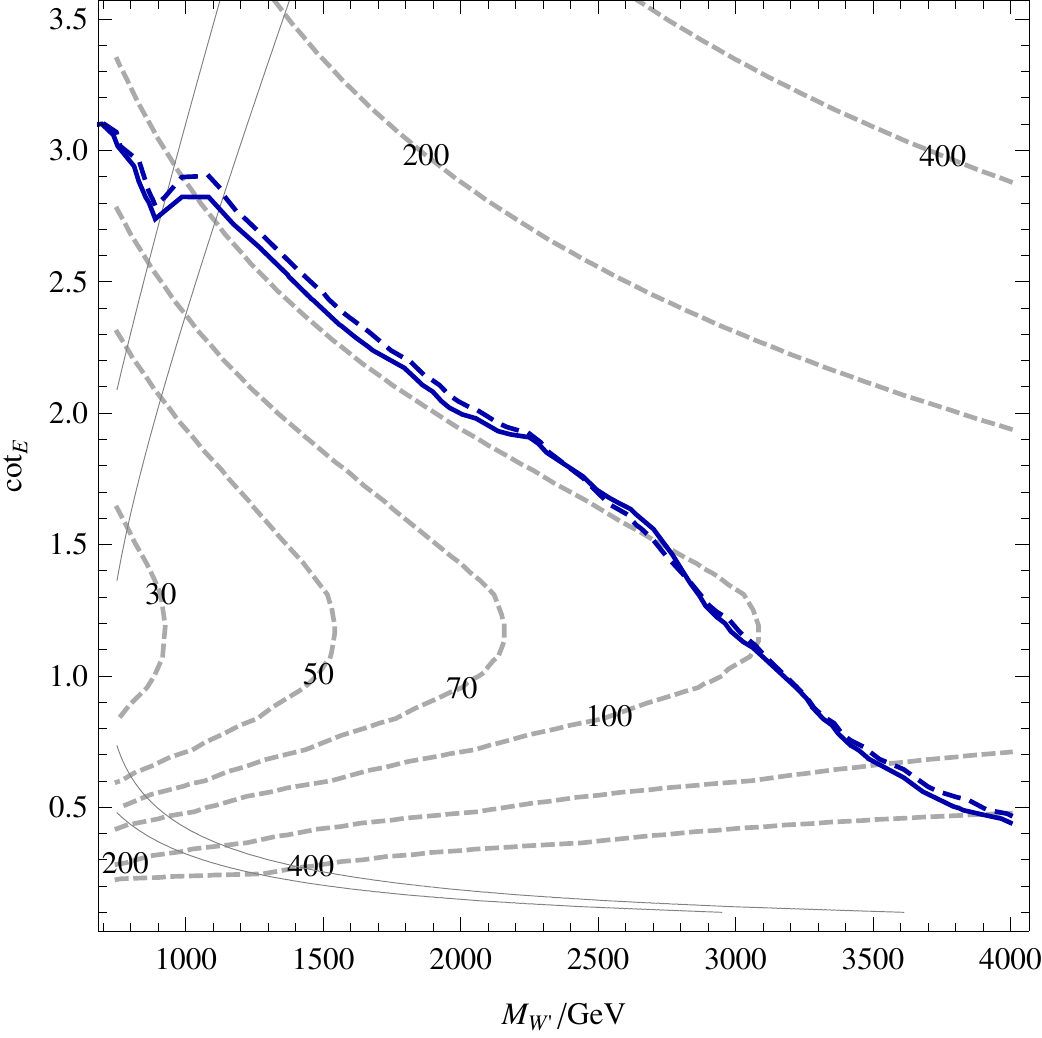}
\caption{Limits from  $Z'\rightarrow \tau\tau$ (left)
\cite{TheATLAScollaboration:2013yha} and $W'\rightarrow e\nu_e$ (right)
\cite{wprimelep}. The blue dashed lines show the limits from naive scaling of
SSM crosssection, the blue solid lines show the limit after applying the
analysis (and correcting for the SSM-acceptance). The gray dashed lines give
the width of the W'/Z' in GeV. The gray solid lines denote the regions where
$\epsilon^2<0.15,0.1$ (from left to right). \label{fig:searches1}} \end{figure}

In searches for sequential SM-like $Z'$ or $W'$, channels involving $\tau$
leptons or top quarks are currently not particularly competitive. As we have
just seen, this might change when family-nonuniversal couplings are considered
such as they appear in the $SU(2)_1\times SU(2)_2$ model which we have briefly
reviewed above (of course, once we relax the restriction to SM-like heavy
vectors, we need information from all available channels anyhow in order to
classify a newly discovered vector boson).  We now extend the discussion of
bounds on this model given in\footnote{Oddly, in \cite{Kim:2011qk,Kim:2014afa},
the authors consider the limit $\tan\beta\rightarrow 0$, which is not
theoretically feasible, especially without completely changing the remaining
phenomenology of the model.  Fortunately, the $V'$ searches are only weakly
affected by this choice.} \cite{Kim:2011qk,Kim:2014afa} by recent searches and
a discussion of model dependence.  Indeed, the other crucial difference
apart from the modified couplings is the modified width of the heavy vectors
which can be significantly enhanced with respect to the sequential case (the
respective widths are shown in Figures \ref{fig:searches2} and
\ref{fig:searches1}). This can render problematic any comparison with existing
searches which are specifically performed or at least optimized for sequential
SM-like vectors, or in general for particles with narrow width. 

For example, a recent CMS search presented in \cite{Chatrchyan:2014koa} for
$W'\rightarrow t b$ decays in lepton+jets, which to date is the most sensitive
analysis for this channel, uses a shape-based Bayesian analysis. While the
quoted limit including interference effects of $M(W')>1.84$ TeV (improving to
$M(W')>2.05$ TeV without interference) can be interpreted in terms of an upper
bound on $BR\times \sigma$, it is not obvious how this bound changes when
realistic widths are taken into account.  Since the analysis is too intricate
to redo, we limit ourselves to naively scaling the bound on $BR\times \sigma$
for the sequential case according to the modified couplings and branching
ratios in the nonuniversal model. The result is shown in Figure
\ref{fig:searches2} (red dashed).  Note also that interference effects will be
influenced by the relative sign of $W$ and $W'$ couplings, which is reversed in
the model we consider here and could lead to a further enhancement. It would
therefore be interesting to know the exclusion including interference effects
of either sign, and for general widths.

For $\cot_E \gtrsim 2$, the limits from $W'\rightarrow t b$ become more
sensitive than the $Z'\rightarrow ll$ search presented in \cite{Aad:2014cka},
which we have similarly scaled to accomodate modified couplings and branching
ratios (see Figure \ref{fig:searches2}, blue dashed).

The CMS search in $Z'\rightarrow t\overline{t}$ \cite{CMS:lhr} in principle
profits from the enhancement of third generation couplings, but is currently
not competitive with other final states in the model under consideration.

Two further searches, $Z'\rightarrow \tau \tau$
\cite{TheATLAScollaboration:2013yha} and $W'\rightarrow e\nu$ \cite{wprimelep},
are based on a single-bin analysis using an $M_T^{min}$ cut. An $M_T$ cut is
chosen depending on the signal mass hypothesis in order to optimize the
expected $S/B$ in the case of a sequential $Z'$ and $W'$. Since the quoted
limits are  in principle sensitive to width effects, we have implemented these
searches into ROOT, and simulate $W'$ and $Z'$ production using MadGraph
5\cite{Alwall:2014hca}/Pythia\cite{Sjostrand:2006za}/Delphes\cite{deFavereau:2013fsa}
using our FeynRules \cite{Christensen:2008py,Alwall:2014bza} implementation of
the $SU(2)_1\times SU(2)_2$ model.  We have then determined the relative
acceptances of the $M_T$ cut in the nonuniversal model compared to the
sequential case, and rescale the limits accordingly. The results (solid) are
shown in Figure \ref{fig:searches1} in comparison with the naively scaled
limits (dashed). We find that in the region of enhanced width, the discrepancy
is $\gtrsim 100$ GeV in the case of the $Z'\rightarrow \tau \tau$ final state,
and as expected, the naively scaled result somewhat overestimates the exclusion
power. The effect in the $Z'\rightarrow l l$ search is less pronounced, which
was to be expected since the sensitivity drops off quickly for $\cot_E>1$, and
the widths in the excluded region therefore stay below $\lesssim 100$ GeV.
While it is probably unavoidable that in the case of increased width, $S/B$
decreases, the $M_T^{min}$ cuts used in these searches could still be optimized
for different widths, which would potentially improve their sensitivity to
non-sequential vectors. 

In fact, a CMS search for $W'\rightarrow \tau \nu$ is in preparation which
takes this effect into account by optimizing the signal mass dependent
$M_T^{min}$ cuts using the widths suggested by the family-$SU(2)_1\times
SU(2)_2$ model. It has the potential to yield the strongest bound in the region
$\cot_E>1$.  \label{sec:searches}

\section{Limits from Higgs Searches}
\label{sec:higgs}
There are two obvious ways in which heavy vectors could contribute to a signal
in ``Higgsstrahlung'' type searches\footnote{While this paper was in
preparation, \cite{Hoffmann:2014aha} came out which uses Higgs searches in
order to constrain composite Higgs scenarios. Repurposing Higgs searches for
$V'$ searches has been suggested for example in
\cite{Li:2013ava,Diaz-Cruz:2013kpa}.}. There could be radiation of a Higgs off
a $V'$, $$pp\rightarrow V^*,{V'}^* \rightarrow V' h\,,$$ followed by suitable
$V'$ decay modes, or resonant $V'$ production with subsequent decay, $$pp
\rightarrow V' \rightarrow Vh\,.$$ The former process is strongly suppressed by
the $V'$ mass, and due to the background-like kinematic configuration (back-to
back $V'$ decay products and a soft Higgs), the Higgs searches which we
consider here are not sensitive to it.  The latter process on the other hand is
equivalent to standard $V'$ production, and in the model under consideration,
merely suffers from a somewhat reduced branching fraction due to the factors
$\frac12$ or $\frac14$ relative to $V'\rightarrow ff$. Furthermore, its
kinematics are identical to the high-$p_T$ tail of conventional SM associated
Higgs production, and acceptance of selection cuts is in principle excellent.
We now concentrate on the search in ATLAS for Higgs associated production in
final states with two $b$-jets, leptons and missing
energy~\cite{TheATLAScollaboration:2013lia} which are designed for production
processes 
\begin{eqnarray} p p &\longrightarrow& Zh; \,\,h\longrightarrow b\overline b, Z \longrightarrow l\overline l \nonumber,\nu \overline \nu \\
p p &\longrightarrow& W^\pm h; \,\,h\longrightarrow b \overline b, W^\pm \longrightarrow \overline l \nu/ l \overline \nu\,.
\end{eqnarray}
This ATLAS search was recently discussed in the context of EFTs precisely for
its sensitivity to high-$p_T$ effects \cite{Biekoetter:2014jwa,Ellis:2014dva}
which we want to exploit here as well.  We have  implemented the corresponding
ATLAS searches~\cite{TheATLAScollaboration:2013lia} into ROOT, and simulations
are again performed using MadGraph/Pythia/Delphes using our FeynRules
implementation of the $SU(2)_1\times SU(2)_2$ model. The analyses of
Ref.~\cite{TheATLAScollaboration:2013lia} use 5 ($2l$ and $1l$) or 3 ($0l$)
different $p_T(V)$ bins separated at $p_T(V)=
(0-90,90-120),120-160,160-200,>200$ GeV which are subject to different
additional kinematic cuts. In the case of heavy vectors decaying like $V'
\rightarrow h+V\rightarrow  \overline b b+ \dots$, virtually all signal events
from onshell production fall into the highest-$p_T$ bin. For $m_{V'}> 400$ GeV,
the search as it is performed is thus not very sensitive to the width of the
heavy vector.  The overall sensitivity is however reduced by only using the
leptonic decay modes of the $W,Z$.  While this is sensible for SM Higgs
production, sensitivity to $V'$ production might be improved in this high-$p_T$
region by using dedicated searches in $\overline b b+jets$ final states, where
one can additionally exploit the fact that $m_{bb}\approx m_h$.  However, let
us press on to see how far we get by using the existing data in the leptonic
final state.  For different values of $\tan_E$, $\sigma\times BR$ scales (up to
PS factors) like $V'$ production and subsequent decay to third generation
fermions  which was discussed before.

\begin{figure}
\begin{center}
\includegraphics[height=7.5cm]{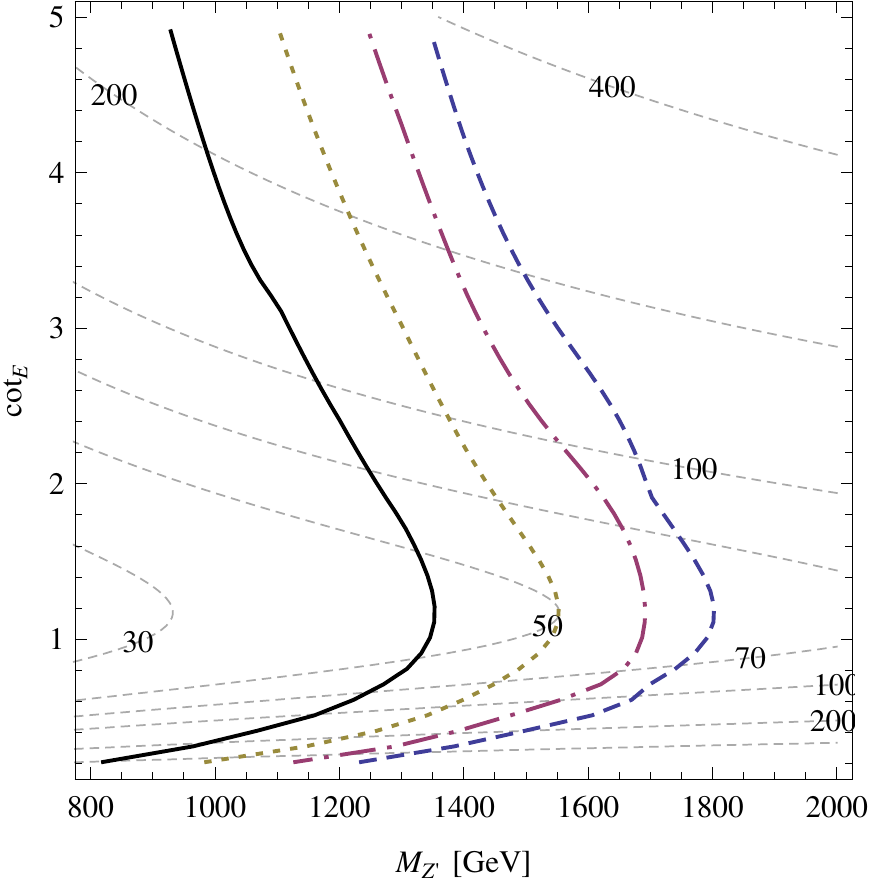}
\caption{
The limit (solid black) on family-$SU(2)_1\times SU(2)_2$ model $W'$ and $Z'$
production at the LHC which we obtain from ATLAS
searches\cite{TheATLAScollaboration:2013lia} for $Vh$ associated production
using 7 and 8 TeV data. The yellow dotted, red dotted-dashed and blue dashed
contours show the parameter values for $10,5,3$ expected events in the 1-lepton
channel respectively. \label{fig:higgsexclusion}}
\end{center}
\end{figure}

The acceptance for the three classes of final states improves for masses
$m_{V'}\gg m_W$, reaching $\approx 25\%$, but deteriorates for $m_{V'}\gtrsim$
TeV due to decreasing angular separation of the two jets originating from a
boosted Higgs.  We use the $70\%$ b-tagging efficiency which is specified in
the ATLAS search throughout the energy range, which may be a source of
additional uncertainty. Furthermore, we add the $2$-jet and $3$-jet tagged
events for simplicity.  Knowing the number of observed events, expected
background, background error and expected SM Higgs signal in the three overflow
bins, we calculate the $\Delta \chi^2$ in the presence of an additional $V'$
signal for a given mass.  Note that we use the SM Higgs signal as a background.
Although there is a slight underfluctuation in the 2-lepton final state, the
best fit point is approximately given for $m_{V'}\rightarrow \infty$. The
resulting limit is shown in Figure \ref{fig:higgsexclusion}.  We have not
included $K$-factors here, which are usually $\gtrsim 1$ for $W',Z'$ production
and could further enhance the sensitivity slightly.  While this limit is not
very strong compared to dedicated $V'$ searches, it is noteworthy that it can
be obtained with existing published Higgs data. Of course, having the signal
events in one overflow bin together with all $p_T^V>200$ background events is
hopelessly conservative for our application. Sensitivity could obviously be
improved by imposing optimized $p_T^V$ cuts for each signal mass hypothesis
$m_{V'}$ in analogy to the variable $M_T^{min}$ cuts used for the $W'$ and $Z'$
searches discussed above. In order to do this, we need to know the $p_T$ shape
of relevant backgrounds after cuts. Of course, by optimizing this $p_T^V$ cut
for maximal expected S/B, we once again gain sensitivity to the total width,
and again, experimental analyses should ideally take this into account in order
to be applicable to a wide range of $V'$ models. To obtain an estimate
what the actual limits on such $V'$ searches are, we have a closer look at the
backgrounds. The 1-lepton channel by itself yields the strongest limits, and
its background is dominated by top production, in particular $\overline t t$
decaying semileptonically. We simulate the process at LO (+radiation of a hard
jet) using MG5/Pythia/Delphes as we did with the signal, and perform the $Vh$
analysis, but with different choices of $p_T^V$ cutoff.  We are particularly
interested in the mass scale where no or few background events are expected.
We find that the background from $t \overline t$ production yields $O(1)$
events for $p_T^V > 400$ GeV.  A naive estimate of significance from Poisson
statistics suggests that for $S\geq 3$ expected events, no observation
corresponds to $p\leq e^{-3} \sim 0.05$. To estimate the potential exclusion
power, in Figure \ref{fig:higgsexclusion} we have marked the parameter regions
with $3$, $5$ and $10$ expected events in the 1-lepton search for $W'$
production.  These numbers will still be subject to the $p_T^V$ search cut.
While the majority of events from the decay will have $p_T^V\sim M_{V'}/2$ and
thus lie in a region with negligible SM+Higgs background, there is a suppressed
tail towards smaller values. However, for a $2 p_T^{V,min}\ll M_{V'}$, only a
small percentage of events are cut away. For example, for $p_T^{V,min}=400$ GeV
and $M_{V'}=1700$ GeV, we find an acceptance after all other cuts of $\mathcal
A(p_T^{V,min})\gtrsim 90\%$.  A complete experimental analysis of the data with
all backgrounds, generalizing the Higgs searches in associated production and
fully exploiting the known invariant mass of Higgs decay products, would be an
interesting endeavour.

The description of BSM effects using 59 dimension-6 operators
\cite{Buchmuller:1985jz} relies on a minimal flavour violation (MFV) scheme. In
this framework, the leading BSM effects affecting only Higgs physics and triple
gauge couplings (TGCs) can be described by (depending on how one counts) 10
operators \cite{Elias-Miro:2013mua,Pomarol:2013zra}. Models with nonuniversal
couplings to fermions such as the one discussed here require an extension of
this basis. Even in the ``universal'' limit $\tan_E=1$, the model predicts a
relative sign between the 3rd generation $V'$ couplings and the others,
spoiling an exact matching to the MFV operator basis. However, we can make a
quantitative comparison as long as we are dealing with couplings of the $V'$ to
1st and 2nd generation fermions, as is the case in this Higgs search. In our
case, the relevant couplings corresponding to the Feynman rules given above are 
\begin{equation}
i\mathcal L = i \frac{g}{\sqrt{2}} \overline u \gamma^\mu P_L d\, {W}^{\prime +}_\mu-~ i \frac12 g^2 \phi^\dagger \phi {W}^{\prime -}_\mu W^{+\mu}
\end{equation}
where $\phi$ denotes the light SM Higgs doublet. Integrating out a $W'$ of mass
$M$ at tree level amounts to putting the propagator $i g^{\mu\nu}/M^2$ between
the two dim-4 operators, resulting in a dimension-6 interaction
\begin{equation}
i\mathcal L_{eff} = i \frac{g^3}{2\sqrt{2} M^2}\, \phi^\dagger \phi\, (\overline u \gamma^\mu P_L d) \, W^+_\mu\,.
\end{equation}
Due to gauge invariance, this interaction can be mapped to a term
\begin{equation}
\mathcal L_{eff} = \frac{g^2}{4 M^2} i (\phi^\dagger \sigma^a \overleftrightarrow D_\mu \phi) \overline Q_L \sigma^a \gamma^\mu Q_L
\label{dim6:fermionic}
\end{equation}
in the Lagrangian. It can be eliminated for the first two generations by
performing a field redefinition 
\begin{equation}
W^a_\mu \longrightarrow W^a_\mu - \frac{g}{2 M^2}\,i \phi^\dagger\sigma^a \overleftrightarrow D_\mu \phi
\end{equation}
which generates
\begin{equation}
\Delta\mathcal L = \frac{1}{M^2} \frac{i g}{2} (\phi^\dagger \sigma^a \overleftrightarrow D_\mu \phi) D_\nu W^{a\mu\nu} \equiv \frac{1}{M^2} \mathcal O_W
\end{equation}
as well as $2\mathcal O_H-4 \mathcal O_r= (\phi^\dagger \sigma^a
\overleftrightarrow D_\mu \phi)^2$ at dimension-6 level\footnote{Note that the
same operator $\mathcal O_H-2 \mathcal O_r$ is already generated upon
integrating out $W'$, $Z'$, but enhanced with $\cot_E^2$. It rescales the Higgs
couplings and becomes an important effect in the $\cot_E\gg 1$ regime. A
detailed analysis of this effect is subject for future research, and will not
concern us for the comparison at hand which focuses on the effects of $\mathcal
O_W$.}, and an additional term of the form Eq. (\ref{dim6:fermionic}) for the
3rd generation which will cancel the effect of $\mathcal O_W$. 
We can thus identifiy the mass of our heavy vectors with the Wilson coefficient
\begin{equation}\frac{1}{M^2}\sim c_W/m_W^2\end{equation} 
in the basis of \cite{Elias-Miro:2013mua} at least as long as we
consider currents of 1st and 2nd generation fermions. This operator contributes
to the $S$ parameter as well as TGCs and Higgs production.  The bounds on the
combination of operators $\mathcal O_W -\mathcal O_B$ in associated Higgs
production were analyzed in \cite{Biekoetter:2014jwa}, and since we have used
the same ATLAS analysis here to obtain bounds on resonant vector production, we
can make a direct comparison. For the point $\cot_E=1$, the combined exclusion
in the unmodified Higgs search is at $M\gtrsim 1300$ GeV. This translates to a
model-dependent limit on the Wilson coefficient of $|c_W|\lesssim 0.0038$,
which is already significantly smaller than the limits obtained from EFT
analyses of $-0.04<c_W<0.01$. Our estimate for the achievable limit (with current
data) in a Higgs search with optimized $p_T^V$ cuts lies closer to $M\gtrsim
1700$ GeV, which corresponds to $|c_W|\lesssim0.0022$.  The existing $W'$
searches for leptonic final states yield $M\gtrsim 3000$ GeV, corresponding to
$|c_W|\lesssim 0.00072$. 

These EFT analysis in  \cite{Biekoetter:2014jwa} considers the
combination $c_W=-c_B$ in order to probe a direction orthogonal to the S
parameter, which might modify the bound somewhat. 
However, the interpretation of our comparison is not obvious for yet a different reason.
Assuming an underlying model of the weakly interacting type discussed in this paper, bounds of the
order of $-0.04<c_W<0.01$ in the EFT can only be obtained by using the EFT
beyond its actual range of validity: near the resonance, dimension-8 operators
e.g. of the type 
\begin{equation}
\mathcal O_{\Box W}= \frac{i g}{2} (\phi^\dagger \sigma^a \overleftrightarrow D_\mu \phi) D^2 D_\nu W^{a\mu\nu}
\end{equation}
are present from the higher orders in the $p^2$ expansion of the propagator. In
order to be self-consistent in the presence of a weakly-interacting UV
completion, the EFT needs to be cut off below the scale $m_W/\sqrt{|c_W|}$. This reduces
its exclusion power to $c_W<0.01$ (observed due to an underfluctuation, 
no lower limit possible), and no
expected limit at all in either direction\cite{Biekoetter:2014jwa}. Wilson coefficients $c_W\sim -0.04,
0.01,$ if interpreted within our model, furthermore correspond to unrealistically low mass
scales of $M_{Z'}~\sim~400\dots 800$ GeV. The need to cut off the EFT at such
low scales is unfortunate, considering that even using the EFT up to the
unitarity bound yields a very conservative limit. In our example, the price of
model-independence is thus still very high with current data.

This
discrepancy between the two approaches is not surprising, as the EFT precisely
throws away the resonant part of the propagator, only working with the constant
offshell tail at dimension-6 level. The two approaches should converge in
sensitivity for very strongly interacting vectors with large widths, for which
the low-energy effective description nearly saturates the perturbative
unitarity bound. We can observe this effect already here. Indeed, in the model
at hand, we can increase $\cot_E>1$, i.e. $\sin_E<1/\sqrt{2}$, which raises the
gauge coupling $g_2$.  Interestingly, while this soon reduces the $\sigma\times
BR(pp\rightarrow V'\rightarrow Vh)$ due to the $\cot_E^{-2}$ suppression of the
$\overline q q$ initiated production, it leaves the corresponding Wilson
coefficient for the dimension-6 coupling in Eq. (\ref{dim6:fermionic}), which
is proportional to $\cot_E \tan_E/M^2$, unchanged.  The sensitivity of the
direct search in this channel is thus reduced, while the sensitivity of the EFT
analysis remains the same. A similar effect can occur in the
strongly-interacting light Higgs (SILH) model \cite{Giudice:2007fh} also
considered in \cite{Biekoetter:2014jwa}  (in particular the case without
composite fermions), where the respective dependences on the strong coupling
$g_*$ and $g_*^{-1}$  of the $W'-W$ mixing and the Higgs coupling cancel.
There, the comparison is more straightforward since it extends to all
generations universally,  and will be subject to future work.  This property of
the EFT can be seen from the perspective of the full theory: the relative
enhancement of the offshell tail due to the increased width compensates the
reduced coupling.  For a simple weakly interacting heavy sector, the EFT
approach is however still very conservative for existing LHC Higgs data.  This
will in principle change with large luminosity when offshell contributions to
distributions can be measured more accurately while associated resonances are
out of reach.  
\section{Conclusions}
\label{sec:conclusions}
Heavy vector resonances are a well-motivated feature of many extensions of the
SM, and various searches for them are being performed at the LHC. While many
searches concentrate on scenarios in which $W'$ and $Z'$ share the same
couplings as their SM  counterparts (the sequential SM case), and relative
widths are assumed to be small at around $3\%$, there exist well-motivated
extensions of the SM which feature $W'$ and $Z'$ with differing properties. As
an example, we have revisited a family-nonuniversal $SU(2)_1\times SU(2)_2$
model\cite{Chiang:2009kb}.  Depending on the mixing angle of the extended gauge
sector $\cot_E=g_2/g_1$, the model features family-nonuniversal couplings of SM
fermions to the $W'$ and $Z'$.  While searches in channels involving third
generation fermions are generally less sensitive in the sequential SM-like
case, this situation changes for $\cot_E>1$ where couplings to the third
generation are enhanced and all others suppressed.  We have therefore
reinterpreted the limits from ATLAS and CMS searches
$W'\rightarrow t b$ \cite{Chatrchyan:2014koa},
$W'\rightarrow vl$ \cite{wprimelep},
$Z'\rightarrow t\overline{t}$ \cite{CMS:lhr},
$Z'\rightarrow l\overline{l}$\cite{Aad:2014cka} and
$Z'\rightarrow \tau\overline{\tau}$ \cite{TheATLAScollaboration:2013yha}
in terms of the family-nonuniversal model. We find that in large regions of
parameter space, searches involving decays to third generation fermions are
indeed competitive.  However, enhanced third generation couplings also lead to
an enhanced width, which complicates a reliable comparison with the published
searches. While we can only give naively scaled limits for the shape-based
analysis in \cite{Chatrchyan:2014koa}, we have implemented the $M_T^{min}$
based single-bin searches in \cite{TheATLAScollaboration:2013yha} and
\cite{wprimelep} in order to determine the acceptance of the $M_T^{min}$ cuts.
We find that width effects can lead to a significant $O(100)$ GeV
overestimation of the exclusion power.  Furthermore, interference effects are
model dependent. For example, \cite{Chatrchyan:2014koa} find a $\sim 200$ GeV weaker
limit when interference effects with single top production are taken into
account. Since the model we consider here features a relative sign between $W$
and $W'$ couplings to the third generation, this effect might potentially
enhance the sensitivity of this search further. In parameter regions of very
large $\cot_E$ where $\overline q q$ initiated production of $Z'$ is strongly
suppressed, it becomes interesting to consider $pp\rightarrow \overline b b Z'$
associated production. We leave the analysis of this channel to future
research.

In the last section, we have repurposed Higgs
searches\cite{TheATLAScollaboration:2013lia} in vector boson associated
production $pp \rightarrow V h \rightarrow \overline b b + 0l,1l,2l$ to obtain
limits on $Z'$ and $W'$ production.  This channel is of particular interest to
us here because it is sensitive to the Higgs sector mixing angle via
$g_{V'Vh}\propto s_E^2-s_{\beta}^2$ even when the heavy Higgs particles are out
of reach of the LHC.  Indeed, for large $\tan\beta>1$, the $V'V h$ couplings
are enhanced along with those of the third generation.  Using the unmodified
Higgs search, we find a lower bound $M_Z'=M_W'>1300$ GeV, corresponding to a
bound on the Wilson coefficient $|c_W|<0.0038$. However, we have argued that
this limit can be improved considerably by employing additional optimized
$p_T^V$ and $m_{bb}$ cuts for each signal mass hypothesis rather than
collecting all high-$p_T$ events in an overflow bin. In the parameter region of
interest, the reconstruction of highly boosted Higgs becomes crucial (see e.g.
\cite{Schlaffer:2014osa} for a recent analysis).

In a recent paper\cite{Biekoetter:2014jwa} it was argued that in the context of
effective field theories, vector boson associated Higgs production
(``Higgsstrahlung'') is particularly sensitive to operators whose effects
strongly grow with energy, such as $\mathcal O_W$, $\mathcal O_B$ in the
notation of \cite{Elias-Miro:2013mua}.  We have shown that in the low-energy
limit, the family-$SU(2)_1\times SU(2)_2$ provides an interesting testing
ground for the effective theory approach, as the effects of $Z'$ and $W'$ can
be matched to $\mathcal O_W$ as well as some anomalous interactions for the
third generation to which this search is not sensitive, thus allowing for a
direct comparison. We find that in regions of parameter space where
$g_1,g_2<1$, i.e. the weakly interacting regime, direct searches are much more
sensitive than the effective theory approach which essentially discards the
resonance.  However, we observe that for $\cot_E > 1$, the sensitivity of the
direct search decreases, while the Wilson coefficient of $\mathcal O_W$ remains
essentially unaffected.

\section*{Acknowledgements}
We thank Simon Knutzen, Klaas Padeken, Eric Madge Pimentel and Francesco Riva
for valuable discussions.

\end{document}